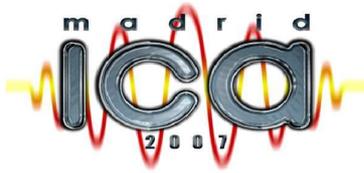

# EXPERIMENTAL VALIDATION OF SIMPLIFIED FREE JET TURBULENCE MODELS APPLIED TO THE VOCAL TRACT

Grandchamp, Xavier; Van Hirtum, Annemie; Pelorson, Xavier
Département Parole et Cognition, GIPSA-Lab, CNRS UMR 5216 INPG ; 46 Av. Félix Viallet, 38031 Grenoble, France; xavier.grandchamp@gipsa-lab.inpg.fr, annemie.vanhirtum@gipsa-lab.inpg.fr, xavier.pelorson@gipsa-lab.inpg.fr

**ABSTRACT**
Sound production due to turbulence is widely shown to be an important phenomenon involved in a.o. fricatives, singing, whispering and speech pathologies. In spite of its relevance turbulent flow is not considered in classical physical speech production models mostly dealing with voiced sound production. The current study presents preliminary results of an experimental validation of simplified turbulence models in order to estimate the time-mean velocity distribution in a free jet downstream of a tube outlet. Aiming a future application in speech production the influence of typical vocal tract shape parameters on the velocity distribution is experimentally and theoretically explored: the tube shape, length and the degree and geometry of the constriction. Simplified theoretical predictions are obtained by applying similarity solutions of the bidimensional boundary layer theory to a plane and circular free jet in still air. The orifice velocity and shape are the main model input quantities. Results are discussed with respect to the upper airways and human sound production.

**INTRODUCTION**
Most research considering the physics of fluid flow in the upper airways deals with the study of voiced sound production or phonation [1,2]. Phonation denotes sounds involving auto-oscillations of the vocal folds due to the interaction with the respiratory airflow. Although the sounds concerned are numerous and can be found in a large portion of world languages, the vocal folds are only a very particular and well localised sound source in the physiological structures situated in the human upper airways. The upper airways consist of the larynx containing the vocal folds and the false vocal folds, the vocal tract and the nasal cavity. Therefore, a major part of fluid flow related phenomena in the upper airways is left untreated in classical studies considering physical phonation modelling. In addition, fluid flow through the upper airways is likely to become turbulent depending on the flow and articulators conditions [3]. The velocity fluctuations in the resulting turbulent upper airway flow as well as the interaction of the turbulent flow with surrounding tissues can act as a sound source for which the created sound power is proportional to a power law of the mean jet velocity [4]. Turbulent flow is associated with several common sounds like aspiration noise and fricatives as well as with abnormal voice conditions as in whispered speech, pathological speech, singing etc. and is furthermore an important actor in the perception of voice quality [3]. Rapid airflow through a constriction is one of the main mechanisms for creating turbulent flow in the upper airways. The airflow at the exit of a constriction is likely to form a jet and turbulent velocity fluctuations are distributed downstream from the constriction exit. The jet characteristics and associated sound production depends on the place, degree and shape of the obstruction. Since the predicted sound power depends on the velocity prediction this paper aims to contribute to the modelling of the mean velocity distribution of turbulent flow through jet formation. In particular the influence of typical vocal tract parameters is explored, i.e. the tube shape, length and the degree and geometry of the constriction. A free jet at a constriction exit is considered. In general three jet regions can be defined, i.e. the mixing, transition and fully turbulent region. Simplified models for the mean velocity distribution based on the similarity principle for subsonic free jets are validated for both the laminar and the turbulent region and a transition criterion is proposed. The model outcome depends on the geometry and the velocity at the exit of the constriction as expressed by the Reynolds number at the constriction exit. Classical phonation models assume a laminar, one-dimensional and steady airflow resulting in severely simplified models derived from the laminar one-dimensional Bernoulli equation corrected for important flow phenomena as flow separation associated with jet formation. In accordance with this successful tradition in physical speech modelling application of laminar losses are discussed. The model outcome is in-vitro validated on extremely simplified geometries for the upstream tube and the constriction since both are approximated by a uniform rectangular or circular shape. In the following the theoretical models are outlined, next the experimental setup is detailed and finally the experimental validation is presented and results are discussed.

**LAMINAR AND TURBULENT FREE JET MODELING**
The Navier-Stokes equations can be severely simplified under certain flow conditions in order to estimate the spatial velocity distribution [5,6,7]. This way the evolution of the mean velocity field for a free jet emerging downstream of either a circular or a uniform constriction for a steady volume flow rate can be modelled from the similarity solutions of the bidimensional boundary layer equations. The following subsections deal respectively with the underlying flow assumptions, the resulting similarity solutions, the transition between the laminar and turbulent flow region and finally in accordance with the tradition in physical vocal folds modelling the application of laminar Bernoulli predicted losses to the jet development.

**Non-dimensional numbers and assumptions**
As for classical vocal folds models several flow assumptions can be formulated from a dimensional analysis of the governing flow equations [8]. The resulting set of non-dimensional numbers serves as a measure of the importance of various flow effects. The development of a free jet at a constriction exit can be characterised from purely geometrical considerations and the velocity at the constriction exit $U_0$. With $h_0$ denoting a typical dimension and $\upsilon$ the cinematic viscosity of air this is expressed by the Reynolds number (Re=$U_0 h_0/\upsilon$). For two dimensional (2D) flow and axisymmetric flow the typical dimension corresponds respectively to two times the outlet constriction half-height $d_0$ or to two times the circular exit radius $r_0$. In the 2D case the flow is assumed to be completely characterised by a bidimensional flow description in the (x,y)-plane where x indicates the flow direction. This implies an aspect ratio $h_0/z_0 \ll 1$, where $z_0$ is a typical dimension perpendicular to the (x,y)-plane. The Reynolds number denotes the importance of inertial forces with respect to viscous forces and provides a flow regime indication, i.e. laminar or turbulent flow. Therefore it is the main feature considered in this study. Various Reynolds numbers will be considered by changing the outlet velocity $U_0$ or the characteristic dimension $h_0$, i.e. $U_0 h_0$, in order to cover different flow regimes. Note that for human speech Re is typically $O(10^3)$. Since steady volume airflow is considered the mean flow is considered to be stationary. Next, the flow is regarded as incompressible since low Mach (Ma) number flow is assumed corresponding to Ma=$U_0/c_0 < 0.4$ where $c_0$ is the speed of sound.

**Bidimensional boundary layer equations and laminar and turbulent similarity solutions**
Applying the above assumptions the Reynolds-averaged Navier-Stokes equations can be reduced to the bidimensional boundary layer approximations. The continuity and momentum equations for laminar and turbulent flows, with the flow index k associated with the radial direction r equal to unity for axisymmetric flows and zero for 2D flows, are given by:

$$\frac{\partial}{\partial x}\left(r^k u\right) + \frac{\partial}{\partial x}\left(r^k v\right) = 0,$$

$$u\frac{\partial u}{\partial x} + v\frac{\partial u}{\partial y} = -\frac{1}{\rho}\frac{dp}{dx} + \frac{v}{r^k}\frac{\partial}{\partial y}\left(r^k \frac{\partial u}{\partial y}\right) + \frac{1}{r^k}\frac{\partial}{\partial y}\left(-r^k \overline{u'v'}\right).$$

where (u,v) and (u',v') denotes respectively the time mean and fluctuating part of the velocity parallel and perpendicular to the flow direction and $\rho$ the density. The Reynolds shear stress term $-\rho\overline{u'v'}$ is equal to zero for laminar flows since the fluctuations are assumed zero. For laminar flows the two remaining unknowns u and v correspond to the searched mean velocity components. The Reynolds stress term in the momentum equation introduces additional unknowns related to the fluctuations for turbulent flows. Consequently additional assumptions are needed in order to reduce the number of unknowns to equal the number of equations and hence to the searched time mean velocity components (u,v). In the current paper the eddy-viscosity approach is applied. This way the bidimensional boundary layer equations for turbulent flow can than be written in the same form as the equations for laminar flow by defining:

$$-\rho\overline{u'v'} = \rho v_t \frac{\partial u}{\partial y}.$$

Moreover the pressure gradient in the free jet emerging from a constriction is small and can be neglected allowing to reduce the momentum equation further, with µ the dynamic viscosity, to:

$$u\frac{\partial u}{\partial x} + v\frac{\partial u}{\partial y} = \frac{1}{\rho r^k}\frac{\partial}{\partial y}\left(r^k \tau\right) \text{ with } \tau = \mu r^k \frac{\partial u}{\partial y} - \rho \overline{u'v'}.$$

The resulting equations admit similarity solutions for the unknown time mean velocity quantities (u,v). This way e.g. $u(x,y)/U_0(x)=F(\eta)$ instead of $u(x,y)/U_0(x)=F(x,y)$ with $\eta(x,y)$ the similarity variable and $U_0(x)$ the centreline velocity along the jet symmetry axis. This way the number of independent variables is reduced from two, i.e. x and y, to one, i.e. $\eta$, and the bidimensionnal boundary layer approximation becomes an ordinary differential equation in u and v. Since the pressure is constant and the motion is steady the total momentum $K=U_0^2 A_0$ in the flow direction is constant with $A_0$ denoting the constriction area. The resulting similarity variables, 2D $\eta(x,y)$ and axisymmetric $\eta(x,r)$, and solutions for the searched time-mean velocities (u,v) are given below for respectively laminar and turbulent jet development. In summary the centreline velocity is proportional to respectively $x^{1/3}$, $x^{-1}$, $x^{-1/2}$ and $x^{-1}$ for the 2D laminar, axisymmetric laminar, 2D turbulent and axisymmetric turbulent jet development.

| 2D laminar jet: | axisymmetric laminar jet: |
|---|---|
| $u = 0.4543 \left(\frac{K^2}{\nu x}\right)^{1/3} \left(1 - \tanh^2 \eta\right)$ <br> $v = 0.5503 \left(\frac{K\nu}{x}\right)^{1/3} \left[2\eta(1 - \tanh^2 \eta) - \tanh \eta\right]$ <br> $\eta = 0.2753 \left(\frac{K}{\nu^2}\right)^{1/3} \frac{y}{x^{2/3}}$ | $u = \frac{3}{8\pi}\left(\frac{K}{\nu x}\right) \frac{1}{(1+\eta^2)^2}$ <br> $v = \frac{1}{2}\sqrt{\frac{3K}{\pi}} \frac{\eta}{x} \frac{1-\eta^2}{(1+\eta^2)^2}$ <br> $\eta = \frac{1}{8}\sqrt{\frac{3K}{\pi}} \frac{1}{\nu} \frac{r}{x}$ |
| 2D turbulent jet: | axisymmetric turbulent jet: |
| $u = \frac{\sqrt{3}}{2}\sqrt{\frac{K\sigma}{x}}\left[1 - \tanh^2(\eta)\right]$ <br> $v = \frac{\sqrt{3}}{4}[2\eta(1 - \tanh^2(\eta)) - \tanh(\eta)]$ <br> $\eta = \frac{\sigma y}{x}$ | $u = \frac{1}{8\alpha}\sqrt{\frac{3K}{\pi}} \frac{1}{x} \frac{1}{(1+\eta^2)^2}$ <br> $v = \frac{1}{2}\sqrt{\frac{3K}{\pi}} \frac{1}{x} \frac{\eta(1-\eta^2)}{(1+\eta^2)^2}$ <br> $\eta = \frac{r}{8\alpha x}$ |

The empirical constants $\sigma$ and $\alpha$ involved in turbulent jet development are both empirical ad-hoc constants determining the eddy viscosity $\nu_t(K,\sigma)$ or $\nu_t(K,\alpha)$ for plane respectively circular jet development, here set to $\sigma=7.67$ and $\alpha=15.2$. Consequently the same way as for the viscosity in a laminar jet, the eddy viscosity is assumed constant in the entire jet and hence independent of $\eta$, but on the contrary the eddy viscosity depends on K. It is easily seen that the presented solutions for the time-mean values of u and v vary as function of $(K,x^n)$ rather than of $(U_0(0),x^n)$, i.e. the total momentum flux K is an input quantity rather than the centreline velocity $U_0(x=0)$ and n determines the jet growth. Since $\rho$ is assumed constant it is omitted as an input variable. The tabled solutions for (u,v) become singular at the origin x=0. Therefore a virtual origin corresponding to a virtual tube length $x_0<0$ needs to be considered. The virtual origin is obtained by assuming that the jet width is tangent to the constriction exit [7]. So with the denominator indicating the jet width or the jet growth $x_0$ for respectively the laminar 2D, laminar axisymmetric, turbulent 2D and turbulent axisymmetric free jet yields respectively:

$$\left(\frac{d_0}{3.203.\left(\frac{\nu^2}{K}\right)^{1/3}}\right)^{3/2} , \quad \frac{r_0}{5.269.\left(\frac{\nu^2}{K}\right)^{1/2}} , \quad \frac{d_0}{0.11} \text{ and } \frac{r_0}{0.086}.$$

Since for turbulent jet growth x0 is independent from K, the spatial jet development is constant and the growth half-angle yields respectively 6.3° and 4.9° for the 2D and axisymmetric jet.

**Laminar to turbulent transition**
A turbulent jet emerging from a constriction is characterised by a laminar near field, a transition region and a fully developed turbulent far field in which the discussed similarity solutions for the time-mean velocity $\sim(x-x_0)^{-n}$ can be applied. In this paper an experimental criteria is proposed to determine the onset of the turbulent far field in case of an axisymmetric turbulent jet growth. The ad-hoc criteria is based on the maximum turbulent Reynolds number defined as $Re_{turb}=u(x,0)\delta(x)/\nu$ where $\delta(x)$ represents the jet width associated with the distance where $u(x,\delta)=u(x,0)/2$ assuming a turbulent jet growth and $u(x,0)$ corresponds to the measured centreline velocity. For the 2D turbulent jet the far field transition is assumed to be reached when $x>12d_0$ in accordance with literature [6].

**Estimation of laminar losses**
Following the tradition in speech production models laminar losses for the retrieved laminar and turbulent development of the jet are assessed by applying the Bernoulli equation to the similarity predicted jet growth. As before the jet width $\delta(x)$ correspond to $u(x,\delta)=u(x,0)/2$. Except for the laminar 2D jet, for which $\delta(x)\sim x^{2/3}$ holds, the jet width is found to be proportional to x, i.e. $\delta(x)\sim x$.

**EXPERIMENTAL SET-UP AND VELOCITY DATA**
In order to validate the presented models for axisymmetric and 2D free jets an experimental setup is used consisting steady flow provided by a valve controlled air supply to which circular and rectangular tubes with different constrictions shapes and sizes can be attached. Five different circular tube lengths where used of 0.03, 0.11, 0.18, 0.20 and 0.5m and two diameters are tested i.e. $2r_0$=0.025m and $2r_0$=0.01m. A plane jet was generated by considering a rectangular constriction of width 0.034m and height $2d_0$=0.018m as well as $2d_0$=0.003m. The volume airflow velocity through the constriction was varied from 20 up to 180l/min in steps of 10l/min corresponding to Reynolds numbers at the constriction exit from $O(10^2)$ up to $O(10^3)$. This allows to study both the laminar and turbulent flow regimes, except for the 2D case where Re~30 is required at the exit in order to generate a laminar flow. This is difficult to realize with the available setup and does not seem to be important aiming speech production applications. A constant temperature anemometer system (TSI4040) is used to measure the downstream mean velocity distribution with an accuracy of 0.1m/s. A sample frequency of 1000Hz is applied and the searched mean velocity magnitude is retrieved by averaging every second. Velocity profiles parallel and perpendicular to the flow direction are obtained by mounting the hot film on a two-dimensional stage positioning system (IFA300). The accuracy of positioning in the x- and y-direction is 4 and 2µm. A step size $\Delta x$=1cm in the parallel and $\Delta y$=1mm in the perpendicular direction is applied. The longitudinal velocity profile is considered from the exit up to 40cm downstream along the jet centreline. Transverse profiles are assessed at 20, 30 and 40cm downstream from the constriction and span +/-8cm across the jet centreline. Longitudinal profiles and transverse profiles are 2 and 4 times repeated resulting in respectively 2 and 4 datasets for the centreline and transverse time mean velocities for each assessed configuration of tube, constriction and volume airflow. In the following all 2 or 4 repetitions are shown.

**RESULTS AND DISCUSSION**
The influence of the upstream tube shape and length as well as the shape and degree of an exit constriction on free plane or circular jet development is experimentally explored for different flow regimes. The outlined model predictions are validated against the experimental observations. The shown velocities are normalised with respect to the measured velocity at the exit ($U_0$) while distances are normalised with respect to the diameter ($2r_0$) or the constriction height ($2d_0$).

**Axisymmetric circular laminar jet**
Measured and modelled longitudinal velocities for Re=565 obtained for a circular exit of 25mm diameter and 5 different duct lengths of 3,11, 18, 20 and 50cm are presented in Figure 1.

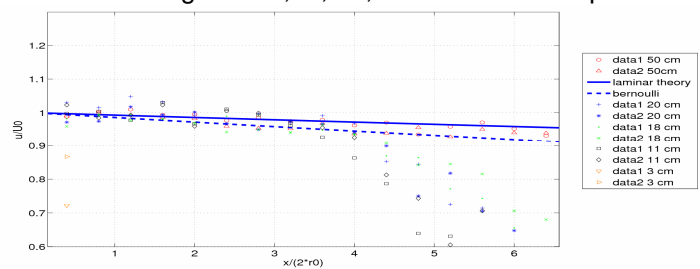

Figure 1 - Axisymmetric longitudinal velocity profiles for Re=565 and different tube lengths.

The laminar similarity solution as well as the Bernoulli laminar loss estimation predicts the time mean velocity in the laminar part of the jet well within 5%. The laminar character following the tube exit ranges up to 6 times the diameter for a 50cm upstream tube, while for a 3cm upstream tube the laminar region spans less than 1 diameter and hence laminar model predictions fails immediately following the tube exit. For intermediate tube lengths the laminar regime extents up to about 4 times the tube diameter. So the spatial extension of the laminar jet regime is related with the upstream tube length. Out of the laminar regime the velocity is severely overestimated by the laminar models.

**Axisymmetric circular turbulent jet**
The influence of the upstream tube length and the virtual origin on the prediction accuracy is further illustrated in Figure 2. Measured and predicted longitudinal velocity profiles are presented for the turbulent flow regime Re=4527 up to 18 times the exit diameter.

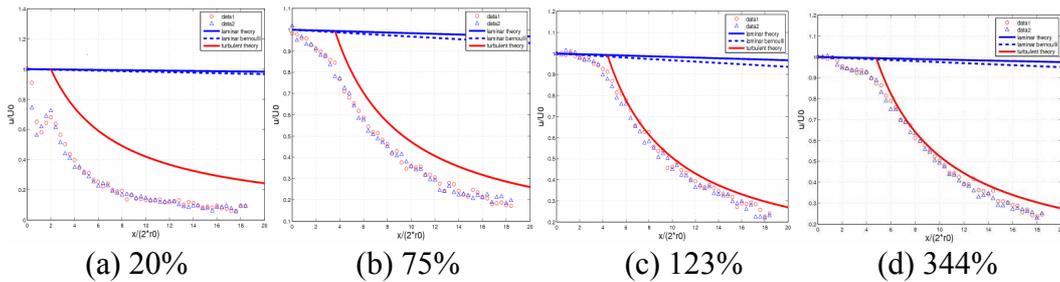

(a) 20%  (b) 75%  (c) 123%  (d) 344%

Figure 2 - Axisymmetric longitudinal velocity profiles for Re=4527 and different tube lengths: (a) 20% of $x_0$, (b) 75% of $x_0$, (c) 123% of $x_0$ and (d) 344% of $x_0$.

Different upstream tube lengths are assessed corresponding to 20%, 75%, 123% and 344% of the virtual origin $x_0$ required in the model. The near, transition and far regions of the turbulent jet development can be identified. As a reference the laminar similarity solution and the laminar Bernoulli approximation are illustrated up to 20 times the exit diameter. As expected laminar models predict the measured time mean velocity data well, i.e. error<5%, in the near field immediately following the constriction exit and rather well, error<10-20% in the transition region, but severely overestimates, >20%, the searched velocities in the far field. In the far field the quantitative behaviour of the losses is described by the turbulent similarity solutions. The qualitative accuracy of the turbulent predictions depends clearly on the upstream tube length and the virtual origin $x_0$. If the upstream tube length contains the virtual origin the prediction accuracy is well within 5% as illustrated in part a and b of Figure 2. The model accuracy decreases when the ratio of the virtual origin and the upstream tube increases. This is illustrated in part a and b of Figure 2 where the overall averaged accuracy in the far field yields about 25% and 15%. The spatial onset of the turbulent transition was determined from the ad-hoc transition criterion considering $Re_{turb}$. Figure 2 illustrates the satisfying performance of the proposed criterion. Note that again it is found both from the measured data as from the predicted transition that the spatial transition onset between the laminar near field and the turbulent far field occurs nearer to the exit for short tube lengths than for long tube lengths and this specifically when the virtual origin exceeds the duct length. The turbulence models use the velocity predicted from the laminar similarity solution at the transition point as an input to calculate the total momentum and hence the jet quantities. Remark that reducing this input velocity is likely to improve the model outcome to a fair extent and this in particular when the upstream tube length is much smaller than the required virtual origin.

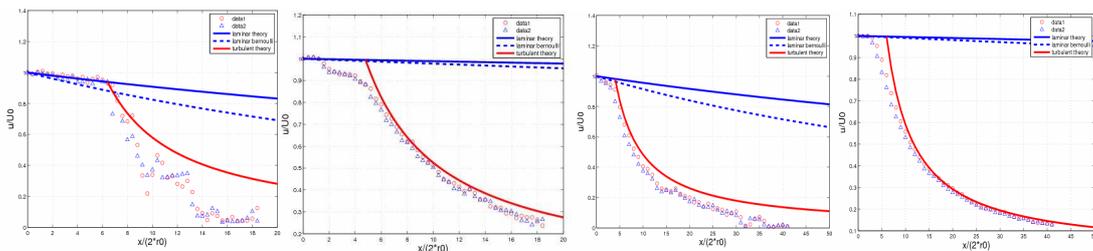

(a) 25mm, 20l/min  (b) 25mm,180l/min  (c) 10mm, 20l/min  (d) 10mm, 180l/min

Figure 3 - Axisymmetric longitudinal velocity profiles tube length equal to 344% of $x_0$ and constriction diameter of 25 and 10mm and volume flow rates of 20 and 180l/min.

Next the influence of the diameter and the volume flow rate on the model accuracy is assessed. Figure 3 illustrates experimental and predicted longitudinal time mean velocities for a circular tube with diameter 25mm and length equal to 344% of $x_0$. Measurements for two constriction diameters, i.e. 25mm and 10mm, and two volume airflow rates, i.e. 20l/min and 180 l/min are shown. Decreasing the constriction diameter or increasing the volume airflow rate raises Re and the momentum K. High Re or K values involves higher velocities, so the longitudinal profiles are shown up to 40 instead of 20 times the constriction diameter for high Re values. The transition to turbulence is predicted to occur about 5 times the diameter regardless the Re number. Both the near as the far field are predicted very accurately within 10% by the laminar and turbulent similarity solutions for the entire turbulent jet development.

**Two-dimensional plane turbulent jet**

Longitudinal velocity profiles up to 160 times the constriction height $2d_0$ at the exit and transverse velocity profiles for the two-dimensional plane turbulent jet are illustrated in Figure 4. The upstream tube length equals 2 times the required virtual origin $x_0$. Again the ad-hoc transition criterion predicting transition to occur at 12 times $d_0$ or 6 times $2d_0$ is in good agreement with the experimental data. Remark that for the axisymmetric turbulent jet 6 times $2r_0$ is the upper limit for transition predicted in case the upstream tube length is larger than the virtual origin $x_0$. Furthermore both the laminar as the turbulent similarity solutions predict the data very accurately up to 60 times the constriction height at the exit with a prediction error inferior to 5%. Hereafter the velocities become very small and the turbulent similarity solution overestimates the measured velocities with 15%. Part c and d of Figure 4 illustrate the good prediction accuracy for the transverse velocity profiles with the worst case error <20%. In general the accuracy of the transverse profile and so the jet width depends on the estimation of the maximum velocity and hence the prediction accuracy of the longitudinal velocity profiles.

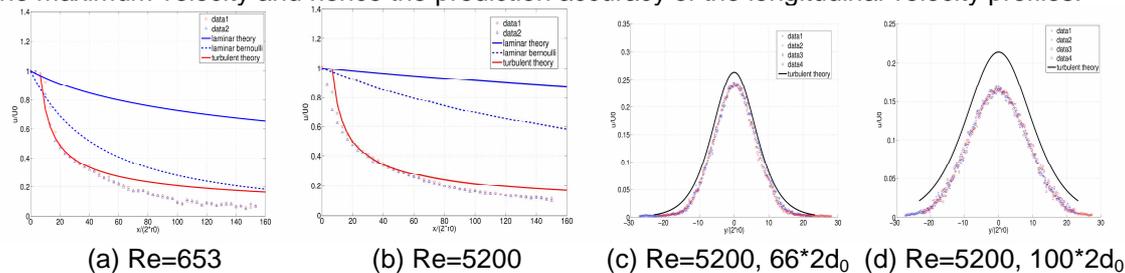

(a) Re=653    (b) Re=5200    (c) Re=5200, 66*$2d_0$    (d) Re=5200, 100*$2d_0$

Figure 4 – 2D longitudinal and transverse velocity profiles for (a) Re=653, (b) Re=5882, (c) Re=5200 at 66 times $2d_0$ and (d) Re=5200 at 100 times $2d_0$.

**CONCLUSION**

Similarity solutions of the time-mean velocities of the boundary layer approximations for laminar and turbulent free jet development are validated for two-dimensional and axisymmetric constriction geometries. The overall performance for both the laminar and turbulent jet region is on averaged within 5% and 15% error range. Both longitudinal and transverse velocity profiles are considered. The importance of the upstream tube length is shown.

**Acknowledgements**
The authors acknowledge the Rhone-Alpes region (France) for financial support.